\documentclass[aps,prb,twocolumn,floatfix,showpacs,unsortedaddress,superscriptaddress]{revtex4}

\usepackage{graphics} \usepackage{graphicx} \usepackage{amssymb}
\usepackage{amsfonts} \usepackage{amsmath} \usepackage{bm}

\begin{document}

\title{Spin coupling in zigzag Wigner crystals}

\author{A. D. Klironomos} \affiliation{Materials Science Division,
  Argonne National Laboratory, Argonne, Illinois 60439, USA}
\affiliation{Department of Physics, The Ohio
  State University, Columbus, Ohio 43210, USA}

\author{J. S. Meyer} \affiliation{Department of Physics, The Ohio
  State University, Columbus, Ohio 43210, USA}

\author{T. Hikihara} \affiliation{Department of Physics, Hokkaido
  University, Sapporo 060-0810, Japan}

\author{K. A. Matveev} \affiliation{Materials Science Division,
  Argonne National Laboratory, Argonne, Illinois 60439, USA}

\date{\today}

\begin{abstract}

  We consider interacting electrons in a quantum wire in the case of a
  shallow confining potential and low electron density.  In a
  certain range of densities, the electrons form a two-row (zigzag)
  Wigner crystal whose spin properties are determined by nearest and
  next-nearest neighbor exchange as well as by three- and
  four-particle ring exchange processes. The phase diagram of the
  resulting zigzag spin chain has regions of complete spin polarization and partial spin polarization in addition to a number of unpolarized phases, including antiferromagnetism and dimer order as well as a novel phase generated by the four-particle ring exchange.

\end{abstract}


\maketitle

\section{Introduction}

The deviations of the conductance from perfect quantization in integer
multiples of $G_0=2e^2/h$ observed in ballistic quantum wires at low
electron densities have generated great experimental and theoretical
interest in recent years.\cite{Thomas,Thomas_1,Cronenwett,Kristensen_1,Kristensen,Kane,Thomas_2,
  Reilly_1,Reilly,Crook,Danneau,Rokhinson,klochan,depicciotto,Berggren,
  Spivak,Tokura,Meir,Matveev,bruus1,flambaum,rejec,bruus,hirose,sushkov,
  seelig,Meir2} These conductance anomalies manifest themselves as
quasi-plateaus in the conductance as a function of gate voltage at
about 0.5 to 0.7 of the conductance quantum $G_0$, depending on the
device. Although most experiments are performed with electrons in GaAs
wires,\cite{Thomas,Thomas_1,Cronenwett,Kristensen_1,Kristensen,
  Kane,Thomas_2,Reilly_1,Reilly,Crook,depicciotto} a similar ``0.7
structure'' was recently observed in devices formed in two-dimensional
hole systems.\cite{Danneau,Rokhinson,klochan} It is widely accepted
that the origin of the quasi-plateau lies in correlation effects, but a complete understanding of this phenomenon remains elusive.

Although some alternative interpretations have been
proposed,\cite{depicciotto,bruus1,seelig} most commonly the
experimental findings are attributed to non-trivial spin properties of
quantum wires.\cite{Thomas,Thomas_2,Reilly,Berggren,Thomas_1,Cronenwett,Kane,
  Reilly_1,flambaum,rejec,bruus,hirose,sushkov,Meir,Tokura,Matveev,
  Meir2,Spivak,Rokhinson,Crook} In a truly one-dimensional geometry
the spin coupling is relatively simple: electron spins are coupled
antiferromagnetically, and the low energy properties of the system are
described by the Luttinger liquid theory. The picture may change
dramatically when transverse displacements of electrons are important and the system becomes quasi-one-dimensional.  In particular, the spontaneous spin polarization of the ground state, which was proposed\cite{Thomas,Thomas_2,Reilly,Berggren,Spivak,Rokhinson,Crook} as a possible origin of the conductance anomalies, is forbidden in one
dimension,\cite{Lieb} but allowed in this case.

The electron system in a quantum wire undergoes a transition from a
one-dimensional to a quasi-one-dimensional state when the energy of
quantization in the confining potential is no longer large compared to
other important energy scales.  In this paper we consider the spin
properties of a quantum wire with shallow confining potential.  In
such a wire the electron system becomes quasi-one-dimensional while
the electron density is still very low, and thus the interactions
between electrons are effectively strong.  At very low densities, electrons in the wire form a one-dimensional structure with
short-range crystalline order---the so-called Wigner crystal.  As the
density increases, strong Coulomb interactions cause deviations from
one-dimensionality creating a quasi-one-dimensional zigzag crystal
with dramatically different spin properties.  In particular, ring
exchanges will be shown to play an essential role.

We find several interesting spin structures in the zigzag crystal.  In
a sufficiently shallow confining potential, in a certain range of electron densities, the 3-particle ring exchange dominates and
leads to a fully spin-polarized ground state.  At higher electron
densities, and/or in a somewhat stronger confining potential, the
4-particle ring exchange becomes important.  We study the phase
diagram of the corresponding spin chain using the method of exact
diagonalization, and find that the 4-particle ring exchange gives
rise to novel phases, including one of partial spin polarization.

The paper is organized as follows.  The formation of a Wigner crystal
in a quantum wire and its evolution into a zigzag chain as a function
of electron density are discussed in Sec.~\ref{WC}. Spin
interactions in a zigzag Wigner crystal which arise through
2-particle as well as ring exchanges are introduced in Sec.~\ref{SE}. The numerical calculation of the relevant exchange constants
is presented in Sec.~\ref{EX}. The results of the numerical
calculation establish the existence of a ferromagnetic phase at
intermediate densities and the dominance of the 4-particle ring
exchange at high densities. Subsequently, a detailed study of the
zigzag chain with 4-particle ring exchange is presented in
Sec.~\ref{4P}. An attempt to construct the phase diagram for a realistic quantum wire as a function of electron density and interaction strength is presented in Sec.~\ref{PDS}. The paper concludes with a discussion of the relation of our work to recent experiments, given in Sec.~\ref{disc}. A brief summary of some of our results has been reported previously in Ref.~\onlinecite{us}.

\section{Wigner Crystals in Quantum Wires}
\label{WC} We consider a long quantum wire in which the electrons are
confined by some smooth potential in the direction transverse to the
wire axis. Assuming a quadratic dispersion and zero temperature, the
kinetic energy of an electron is typically of the order of the Fermi
energy $E_F=(\pi\hbar n)^2/8m$, whereas the Coulomb interaction energy
is of the order of $e^2n/\epsilon$. Here, $n$ is the (one-dimensional)
density of electrons, $\epsilon$ is the dielectric constant of the
host material, and $m$ is the effective electron mass. As the density
of electrons is lowered, Coulomb interactions become increasingly more
important, and at $n\ll a_B^{-1}$ they dominate over the kinetic
energy, where the Bohr radius is given as
$a_{B}=\hbar^{2}\epsilon/me^{2}$. (In GaAs its value is approximately
$a_{B}\approx\!100${\AA}.)

\begin{figure}[!t]
 \resizebox{.48\textwidth}{!}{\includegraphics{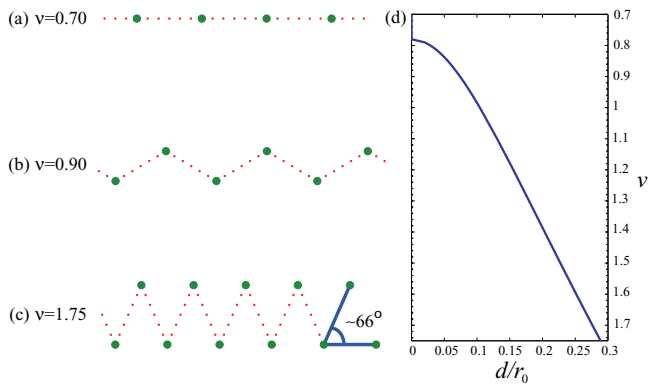}}
  \caption{Wigner crystal of electrons in a quantum wire. The
    structure as determined by the dimensionless distance between rows
    $d/r_0$ depends on the parameter $\nu$ proportional to electron
    density (see text). As density grows, the one-dimensional crystal (a)
    gives way to a zigzag chain (b,c).} \label{fig_1}
\end{figure}

In this low-density limit, the electrons can be treated as classical
particles.  They will minimize their mutual Coulomb repulsion by occupying
equidistant positions along the wire, forming a structure with short-range
crystalline order---the so-called Wigner crystal, Fig.~\ref{fig_1}(a).
Unlike in higher dimensions, the long-range order in a one-dimensional
Wigner crystal is smeared by quantum fluctuations, and only weak density
correlations remain at large distances.\cite{Schulz} However, as will be
shown in the following sections, the coupling of electron spins is
controlled by electron interactions at distances of order $1/n$, where the
picture of a one-dimensional Wigner crystal is applicable. Henceforth, we
speak of a Wigner crystal in a quantum wire with this important
distinction in mind.
Upon increasing the density, the inter-electron distance diminishes,
and the resulting stronger electron repulsion will eventually overcome
the confining potential $V_{\rm conf}$, transforming the classical
one-dimensional Wigner crystal into a staggered or zigzag
chain\cite{hasse,Piacente}, as depicted in Fig.~\ref{fig_1}(b,c). From the comparison of the Coulomb interaction energy $V_{\rm int}(r)=e^{2}/\epsilon r$ with the confining potential an important
characteristic length scale emerges. Indeed, the transition from the
one-dimensional Wigner crystal to the zigzag chain is expected to take
place when distances between electrons are of the order of the scale
$r_0$ such that $V_{\rm conf}(r_{0})= V_{\rm int}(r_{0})$.

It is therefore necessary to identify the electron equilibrium
configuration as a function of density.  In order to proceed in a
quantitative way we consider a specific model, namely a quantum wire
with a parabolic confining potential $V_{\rm
  conf}(y)=m\Omega^{2}y^{2}/2$, where $\Omega$ is the frequency of
harmonic oscillations in the potential $V_{\rm conf}(y)$. Within that
model the characteristic length scale $r_0$ is given as
\begin{equation}
  r_{0}=\left(2e^2/\epsilon m\Omega^2\right)^{1/3}.
\end{equation}
It is convenient for the following discussion to measure lengths in
units of $r_0$. To that respect we introduce a dimensionless density
\begin{equation}
  \nu=nr_{0}.
\end{equation}
Then minimization of the energy with respect to the electron
configuration\cite{hasse,Piacente} reveals that a one-dimensional crystal is stable for densities $\nu<0.78$, whereas a zigzag chain forms at intermediate densities $0.78<\nu<1.75$. (If density is further increased, structures with larger numbers of rows appear.\cite{hasse,Piacente}) The distance $d$ between rows grows with
density as shown in Fig.~\ref{fig_1}. Note that at $\nu\approx 1.46$ the equilateral configuration
is achieved. Therefore, at higher densities---and in a curious contradiction in terms---the distance between next-nearest neighbors is smaller than the distance between nearest neighbors (see Fig.~\ref{fig_1}(c)).

\section{Spin Exchange}
\label{SE} In order to introduce spin interactions in the Wigner
crystal, it is necessary to go beyond the classical limit. In quantum
mechanics spin interactions arise due to exchange processes in
which electrons switch positions by tunneling through the potential
barrier that separates them. The tunneling barrier is created by the
exchanging particles as well as all other electrons in the wire.  The
resulting exchange energy is exponentially small compared to the Fermi
energy $E_{F}$. Furthermore, as a result of the exponential decay of
the tunneling amplitude with distance, only nearest neighbor exchange
is relevant in a one-dimensional crystal. Thus, the one-dimensional
crystal is described by the Heisenberg Hamiltonian $H_1=\sum_j J_1{\bf
  S}_j{\bf S}_{j+1}$, where the exchange constant $J_1$ is positive
and has been studied in detail
recently.\cite{hausler,Matveev,KRM,Fogler} The exchange constant being
positive leads to a spin-singlet ground state with quasi-long-range antiferromagnetic order, in accordance with the Lieb-Mattis theorem.\cite{Lieb}

The zigzag chain introduced in the previous section displays much
richer spin physics. As the distance between the two rows increases as
a function of density, the distance between next-nearest neighbors
becomes comparable to and eventually even smaller than the distance
between nearest neighbors, as illustrated in Fig.~\ref{fig_1}(b,c).
Consequently, the next-nearest neighbor exchange constant $J_{2}$ may
be comparable to or larger than the nearest neighbor exchange constant
$J_{1}$.  Drawing intuition from studies of the two-dimensional Wigner
crystal,\cite{Roger,Katano,Voelker,Bernu} one comes to a further
important realization regarding the physics of the zigzag chain: in
addition to 2-particle exchange processes, \emph{ring exchange}
processes, in which three or more particles exchange positions in a
cyclic fashion, have to be considered in this geometry.

It has long been established that, due to symmetry properties of the
ground state wavefunctions, ring exchanges of an even number of
fermions favor antiferromagnetism, while those of an odd number of
fermions favor ferromagnetism.\cite{Thouless} In a zigzag chain, the
Hamiltonian reads
\begin{eqnarray}
  H&\!\!=\!\!&\frac12\sum_j\Big( J_1 P_{j\,j\!+\!1}+J_2
  P_{j\,j\!+\!2}-J_3(P_{j\,j\!+\!1\,j\!+\!2}
  +P_{j\!+\!2\,j\!+\!1\,j})\nonumber\\
  &&\quad\quad\quad +
  J_4(P_{j\,j\!+\!1\,j\!+\!3\,j\!+\!2}+P_{j\!+\!2\,j\!+\!3\,j\!+\!1\,j})
  -\dots\Big),
  \label{eq:H}
\end{eqnarray}
where $P_{j_1\dots j_l}$ denotes the cyclic permutation operator of
$l$ spins. Here the exchange constants are defined such that all
$J_l>0$. Furthermore, only the dominant $l$-particle exchanges are
shown. A more familiar form of the Hamiltonian in terms of spin
operators is obtained by noting that $P_{ij}=\frac{1}{2} +2{\bf S}_i{\bf
  S}_j$ and $P_{j_1\dots j_l}=P_{j_1j_2}P_{j_2j_3}\dots
P_{j_{l-1}j_l}$.\cite{Thouless}

Using spin operators and considering the two-spin exchanges one
obtains the Hamiltonian
\begin{equation}
  H_{12}=\sum_j\left(J_1{\bf S}_j{\bf S}_{j+1}+J_2{\bf S}_j{\bf
      S}_{j+2}\right). \label{eq-H_12}
\end{equation}
The competition between the nearest neighbor and next-nearest neighbor
exchanges becomes the source of frustration of the antiferromagnetic
spin order and eventually leads to a gapped dimerized ground state at
$J_{2}>0.24J_{1}$.\cite{Majumdar,Haldane,Okamoto,Eggert}

The simplest ring exchange involves three particles and is therefore
ferromagnetic. Including the 3-particle ring exchange $J_3$ in
addition to the 2-particle exchanges, the Hamiltonian of the
corresponding spin chain retains a simple form. The 3-particle
ring exchange does not introduce a new type of coupling, but rather
modifies the 2-particle exchange constants.\cite{Thouless} For a
zigzag crystal we find the effective 2-particle exchange constants
\begin{align}
  \widetilde{J}_{1}&=J_{1}-2J_{3},\\
  \widetilde{J}_{2}&=J_{2}-J_{3}.
\end{align}
Thus the total Hamiltonian has the form
\begin{equation}
  H_{123}=\sum_j\left(\widetilde{J}_1{\bf S}_j{\bf S}_{j+1}
    +\widetilde{J}_2{\bf S}_j{\bf S}_{j+2}\right),\label{new_eq-H_12}
\end{equation}
where $\widetilde{J}_1$ and $\widetilde{J}_2$ can have either sign.

\begin{figure}[!t]
  \includegraphics[height=7.2cm,clip]{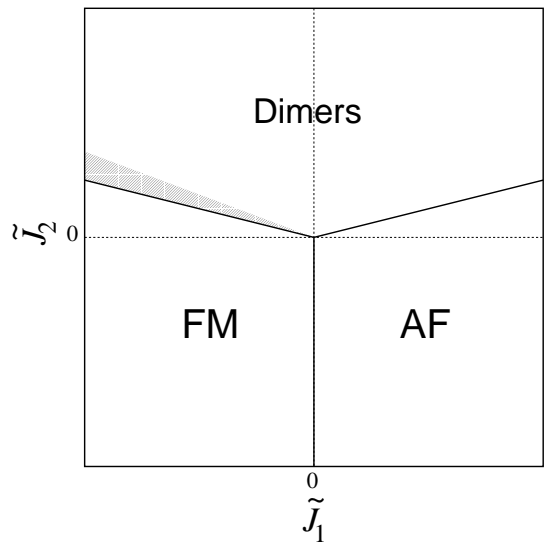}
  \caption{The phase diagram including nearest neighbor, next-nearest
    neighbor, and 3-particle ring exchanges. The effective
    couplings $\widetilde{J}_1$ and $\widetilde{J}_2$ are defined in
    the text. The shaded region between the dimer and ferromagnetic
    phases corresponds to the exotic phase predicted in
    Ref.~\onlinecite{Chubukov}.}
  \label{fig_2}
\end{figure}

Consequently, regions of negative (i.e.~ferromagnetic) nearest and/or
next-nearest neighbor coupling become accessible. The phase diagram of
the Heisenberg spin chain (\ref{new_eq-H_12}) with both positive and
negative couplings has been studied
extensively.\cite{Majumdar,Haldane,Okamoto,Eggert,
  White,Hamada,Tonegawa,Chubukov,Allen,Itoi} In addition to the
antiferromagnetic and dimer phases discussed earlier, a ferromagnetic
phase exists for
$\widetilde{J}_{1}<\min\{0,-4\widetilde{J}_{2}\}$.\cite{Hamada} An
exotic phase called the chiral-biaxial-nematic phase has been
predicted\cite{Chubukov} to appear for $\widetilde{J}_{1} < 0$ and
$-0.25 < \widetilde{J}_{2}/\widetilde{J}_{1} < -0.38$. However, the
nature of the system in this parameter region is still
controversial. The phase diagram is drawn in Fig.~\ref{fig_2}.

Thus, depending on the relative magnitudes of the various exchange
constants, different phases are realized. Extensive studies of the
two-dimensional Wigner crystal have shown that, at low densities (or
strong interactions), the 3-particle ring exchange dominates over
the 2-particle exchanges. As a result, the two-dimensional Wigner
crystal becomes ferromagnetic at sufficiently strong
interactions.\cite{Roger,Bernu} Given that the electrons in a
two-dimensional Wigner crystal form a triangular lattice, by analogy,
one should expect a similar effect in the zigzag chain at densities
where the electrons form approximately equilateral triangles. More
specifically, upon increasing the density and consequently the
distance between rows, one would expect the system to undergo a phase
transition from an antiferromagnetic to a ferromagnetic phase. To
establish this scenario conclusively, the various exchange energies in
the zigzag crystal have to be determined. The system differs from the
two-dimensional crystal in two important aspects. (i) The electrons
are subject to a confining potential as opposed to the flat background
in the two-dimensional case. Even more importantly, (ii) the electron
configuration depends on density, cf.~Fig.~\ref{fig_1}, as opposed to
the ideal triangular lattice in two dimensions. In the following
section, we proceed with a numerical study of the exchange energies
for the specific configurations of the zigzag Wigner crystal in a
parabolic confining potential.

\section{Semiclassical evaluation of the exchange constants}
\label{EX} The effective strength of interactions is usually described
by the interaction parameter $r_s$ which measures the relative
magnitude of the interaction energy and the kinetic energy and is of
order the distance between electrons measured in units of the Bohr
radius. For quantum wires, it is more appropriate to use the parameter
$r_\Omega=r_0/a_B$, which takes into account the confining
potential. Within our model, the interaction parameter $r_\Omega$ is
\begin{equation}
  \label{eq:r_Omega}
  r_{\Omega}=2\left(
    \frac{me^4}{2\epsilon^2\hbar^2}\,\frac{1}{\hbar\Omega}\right)^{2/3}.
\end{equation}
For $r_{\Omega}\gg 1$, strong interactions dominate the physics of the
system, and a semiclassical description is applicable. In order to
calculate the various exchange constants, we use the standard
instanton method, successfully employed in the study of the
two-dimensional\cite{Roger,Voelker,Katano} and
one-dimensional\cite{KRM,Fogler} Wigner crystal. Within this approach,
the exchange constants are given by
$J_{l}=J^{*}_{l}\exp{(-S_{l}/\hbar)}$. Here $S_{l}$ is the value of
the Euclidean (imaginary time) action, evaluated along the classical
exchange path. By measuring length and time in units of $r_0$ and
$T=\sqrt2/\Omega$, respectively, the action $S[\{{\bf r}_j(\tau)\}] $
can be rewritten in the form $S=\hbar\eta\sqrt{r_\Omega}$, where the
functional
\begin{equation}
  \eta[\{{\bf r}_j(\tau)\}]=
  \!\int\limits_{-\infty}^\infty\!\!\mathrm{d}\tau \! \left[\sum_{j}
    \left(\frac{\dot{{\bf r}}_j^2}{2}+y_j^2\right)
    +\sum_{j<i}\frac{1}{|{\bf r}_j\!-\!{\bf r}_i|} \right]\!\!
  \label{eq:action}
\end{equation}
is dimensionless.

Thus, we find the exchange constants in the form
\begin{equation}
  J_{l}=J^{*}_{l}\exp{(-\eta_{l}\sqrt{r_\Omega})}, \label{exch}
\end{equation}
where the dimensionless coefficients $\eta_l$ depend only on the
electron configuration (cf.~Fig.~\ref{fig_1}) or, equivalently, on the
density $\nu$. The exponents $\eta_{l}$ are calculated numerically for
each type of exchange by minimizing the action (\ref{eq:action}) with
respect to the instanton trajectories of the exchanging
electrons. This procedure is mathematically equivalent to solving a
set of coupled, second order in the imaginary time $\tau$,
differential equations for the trajectories ${\bf r}_j(\tau)$. The
boundary conditions at $\tau=\pm\infty$ are, respectively, the
original equilibrium configuration and the configuration where the
electrons have exchanged positions according to the exchange process
considered.
\begin{figure}[!t]
  \includegraphics[height=6.5cm,clip]{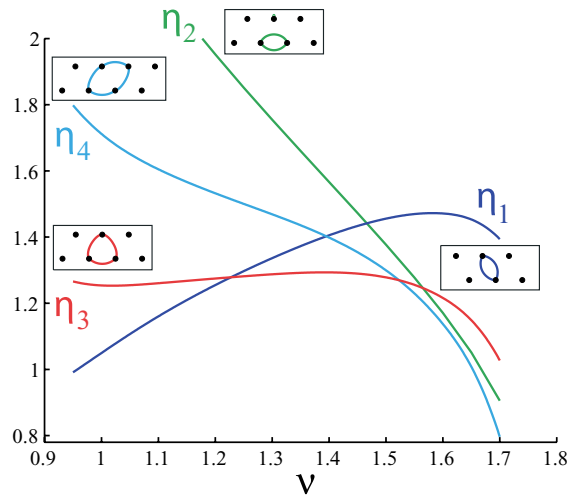}
  \caption{The exponents $\eta_{1}$, $\eta_{2}$, $\eta_{3}$, and
    $\eta_{4}$ as functions of the dimensionless density $\nu$.}
  \label{fig_3}
\end{figure}

In the simplest approximation only the exchanging electrons are
included in the calculation while all other electrons, being frozen in
place, create the background potential. It turns out, however, that it
is important to take into account the motion of ``spectators''---the
electrons in the crystal to the left and to the right of the
exchanging particles---during the exchange process. The results
presented here are obtained by successively adding more spectators on
both sides until the values $\eta_l$ converge. We find that including
12 moving spectators on either side of the exchanging particles
determines the exponents to an accuracy better than $0.1\%$.
\begin{table}[!t]
  \begin{ruledtabular}
    \begin{tabular}{ccccc}
      $\nu$ & $\eta_{1}$ & $\eta_{2}$ & $\eta_{3}$ & $\eta_{4}$\\\hline
      1.0  & 1.050 & 2.427 & 1.254 & 1.712\\
      1.1  & 1.161 & 2.169 & 1.261 & 1.605\\
      1.2  & 1.255 & 1.952 & 1.275 & 1.532\\
      1.3  & 1.337 & 1.754 & 1.287 & 1.469\\
      1.4  & 1.406 & 1.566 & 1.293 & 1.398\\
      1.5  & 1.456 & 1.376 & 1.278 & 1.299\\
      1.6  & 1.471 & 1.169 & 1.215 & 1.135\\
      1.7  & 1.391 & 0.901 & 1.022 & 0.784\\
    \end{tabular}
  \end{ruledtabular}
  \caption{\label{tbl_0}The numerically calculated values of the
    density dependent exponents $\eta_l$, see Eq.~(\ref{exch}). The
    computation was carried out including 12 moving spectator
    particles on either side of the exchanging particles. Corrections
    to all $\eta_{l}$ from the remaining spectators do not exceed
    $0.1\%$.}
\end{table}

Figure \ref{fig_3} shows the calculated exponents for various exchanges as a function of dimensionless density $\nu$ and the corresponding values are reported in Table \ref{tbl_0}. At strong interactions
($r_{\Omega}\gg1$), the exchange with the smallest value of $\eta_{l}$ is clearly dominant, and the prefactor $J^{*}_{l}$ is of secondary importance to our argument.  At low densities, when the zigzag chain is still close to one-dimensional, $J_1$ is the largest exchange constant, and the spin physics is controlled by the nearest-neighbor exchange.  In an intermediate density regime, when the electron configuration is close to equilateral triangles, the 3-particle ring exchange dominates.  Thus, the numerical calculation confirms our original expectation, and a transition from an antiferromagnetic to a ferromagnetic state takes place upon increasing the density.  Surprisingly, however, at even higher densities the 4-particle ring exchange is the dominant process.  The role of the 4-particle ring exchange and the phase diagram of the associated zigzag spin chain will be the subject of the following section. More complicated exchanges have also been computed, namely multi-particle ($l\geq 5$) ring exchanges as well as exchanges involving more distant neighbors.  However, the exchanges displayed in Fig.~\ref{fig_3} were found to be the dominant ones.\cite{us}

It is important to note here that spectators contribute to our results
in an essential way. Allowing spectators to move results not only in
quantitative changes (namely a reduction of the initially
overestimated values $\eta_l$) but in qualitative changes as well: at
high densities, the dominance of the 4-particle ring exchange
$J_{4}$ over the next-nearest neighbor exchange $J_{2}$ is obtained
only if spectators are taken into account. In particular, it is
necessary to include at least 6 moving spectators on each side of the
exchanging particles for $J_{4}$ to take over at high densities.
\begin{figure}[!t]
  \includegraphics[height=4.2cm,clip]{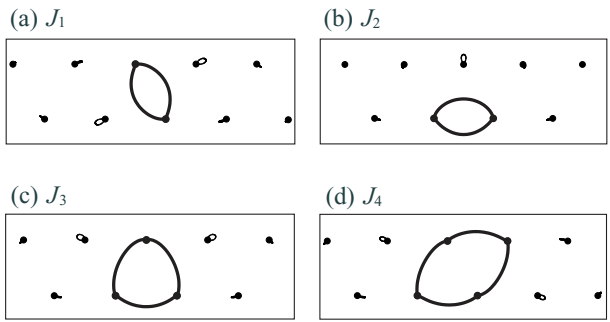}
  \caption{The calculated particle trajectories for various exchanges
    at a representative density $\nu=1.5$. It is evident that only a
    few near neighbors of the exchanging particles move appreciably.}
  \label{fig_4}
\end{figure}

The considerable effect that the spectators have on the values of the
exponents raises the question whether a short-ranged interaction
potential might cause further quantitative or qualitative changes to
the physical picture. In order to investigate that possibility we have
repeated the entire calculation for a modified Coulomb interaction of
the form
\begin{equation}
  V(x) = \frac{e^2}{\epsilon} \left(\frac{1}{|x|} -
    \frac{1}{\sqrt{x^2+(2d)^2}}\right).
\end{equation}
This particular interaction accounts for the presence of a metal gate,
modeled by a conducting plane at a distance $d$ from the crystal. The
gate screens the bare Coulomb potential, modifying the
electron-electron interaction at long distances. Our calculation shows
that this modification affects the values of the exponents only
weakly, even when the gate is placed at a distance from the crystal
comparable to the inter-particle spacing. Qualitatively, the physical
picture remains the same, with the order of dominance of the various
exchanges unaffected throughout the range of densities.

At the same time, it is particularly noteworthy that (both for the
screened and unscreened interaction) the contribution of the spectator
electrons saturates rapidly as their number is increased.  This is an
indication that the destruction of \emph{long-range order\/} in the
quasi-one-dimensional Wigner crystal by quantum fluctuations will not
affect our conclusions.  Figure \ref{fig_4} shows the particle
trajectories for the dominant exchanges at a representative density of
$\nu=1.5$. The trajectories of both the exchanging particles and a
subset of the spectators are shown, and their relative displacements
can be readily compared.

\section{Four-particle ring exchange}
\label{4P}

We have shown in the preceding section that in a certain range of
densities, the 4-particle ring exchange dominates.
Unlike the 3-particle exchange, the 4-particle ring exchange
not only modifies the nearest and next-nearest neighbor exchange
constants, but, in addition, introduces more complicated spin
interactions.\cite{Thouless} For the zigzag chain, we find
\begin{eqnarray}
  \!\!H_4&\!\!=\!\!&J_4\sum_j\Big(\sum_{l=1}^3\frac{4-l}{2} {\bf S}_j{\bf
    S}_{j\!+\!l}+2\big[({\bf S}_j{\bf S}_{j\!+\!1})({\bf S}_{j\!+\!2}
  {\bf S}_{j\!+\!3})\nonumber\\
  &&+({\bf S}_j{\bf S}_{j\!+\!2})({\bf S}_{j\!+\!1} {\bf
    S}_{j\!+\!3})-({\bf S}_j{\bf S}_{j\!+\!3})({\bf S}_{j\!+\!1} {\bf
    S}_{j\!+\!2})\big]\Big). \label{eq-H4}
\end{eqnarray}
Not much is known about the physics of zigzag spin chains with
interactions of this type. We have studied this particular system
described by the Hamiltonian $H=H_{123}+H_4$ using exact
diagonalization, considering systems of $N=12,16,20,24$
sites. Periodic boundary conditions have been imposed, and we have
employed the well-known Lanczos algorithm to calculate a few low-energy eigenstates.

\begin{figure}[!t]
  \includegraphics[height=7cm,clip]{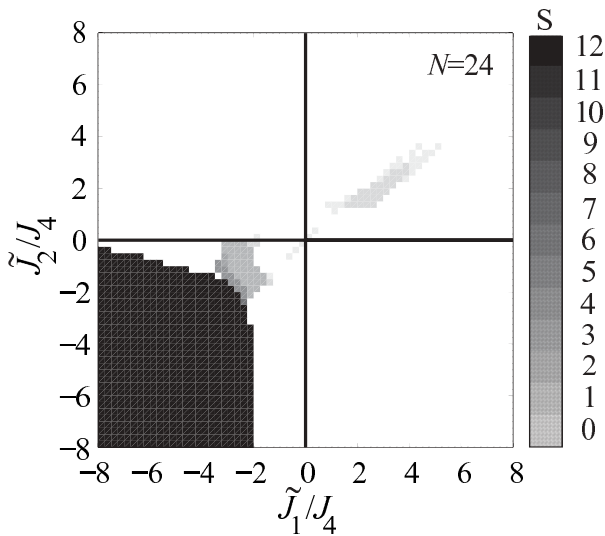}
  \caption{Total spin $S$ of the ground state for a chain of $N = 24$
    sites as a function of the effective couplings
    $\widetilde{J}_1/J_4$ and $\widetilde{J}_2/J_4$.}
  \label{fig_5}
\end{figure}
Figure \ref{fig_5} shows the total spin $S$ of the ground state as a function of the effective couplings $\widetilde{J}_1/J_4$ and $\widetilde{J}_2/J_4$ for the largest system considered, one with $N=24$ sites. The darkest region corresponding to maximal total spin is the ferromagnetic phase, which occurs for large negative couplings in direct analogy to the phase diagram for the system without four-spin interactions (see Fig.~\ref{fig_2}). For all system sizes that we have considered, the obtained phase boundary is almost independent of the system size and agrees very well with the conditions for ferromagnetism
\begin{align}
  \widetilde{J}_1+2J_4&<0,\\
  \widetilde{J}_1+4\widetilde{J}_2+10J_4&<0,
\end{align}
derived by treating the four-spin terms in the Hamiltonian
(\ref{eq-H4}) on a mean field level near the ferromagnetic state.

A new phase of partial spin polarization appears adjacent to the
ferromagnetic phase. The partially polarized phase possesses a ground
state total spin of $S=2$ for $N=12$, $S=2$ or 4 for $N=16, 20$, and
$S=4$ for $N=24$; it appears that total spin of one third of the
saturated magnetization $N/2$ prevails throughout most of that
phase. The phase persists, to a significant extent, in range and form
as $N$ increases. Therefore, we believe it is not a finite size
effect. We note here that it has been shown rigorously that a model
described by a Hamiltonian having a similar form to ours also exhibits
a ground state with partial spin polarization.\cite{Muramoto} On the
other hand, the scattered points corresponding to non-zero total spin
in the first quadrant ($\widetilde{J}_1,\widetilde{J}_2>0$) appear to
shift position as $N$ increases and the size of the total spin remains
small, $S \le 2$, for all system sizes considered. We cannot ascertain
at this point whether they persist in a larger system.

\begin{figure}[!t]
  \includegraphics[height=6cm,clip]{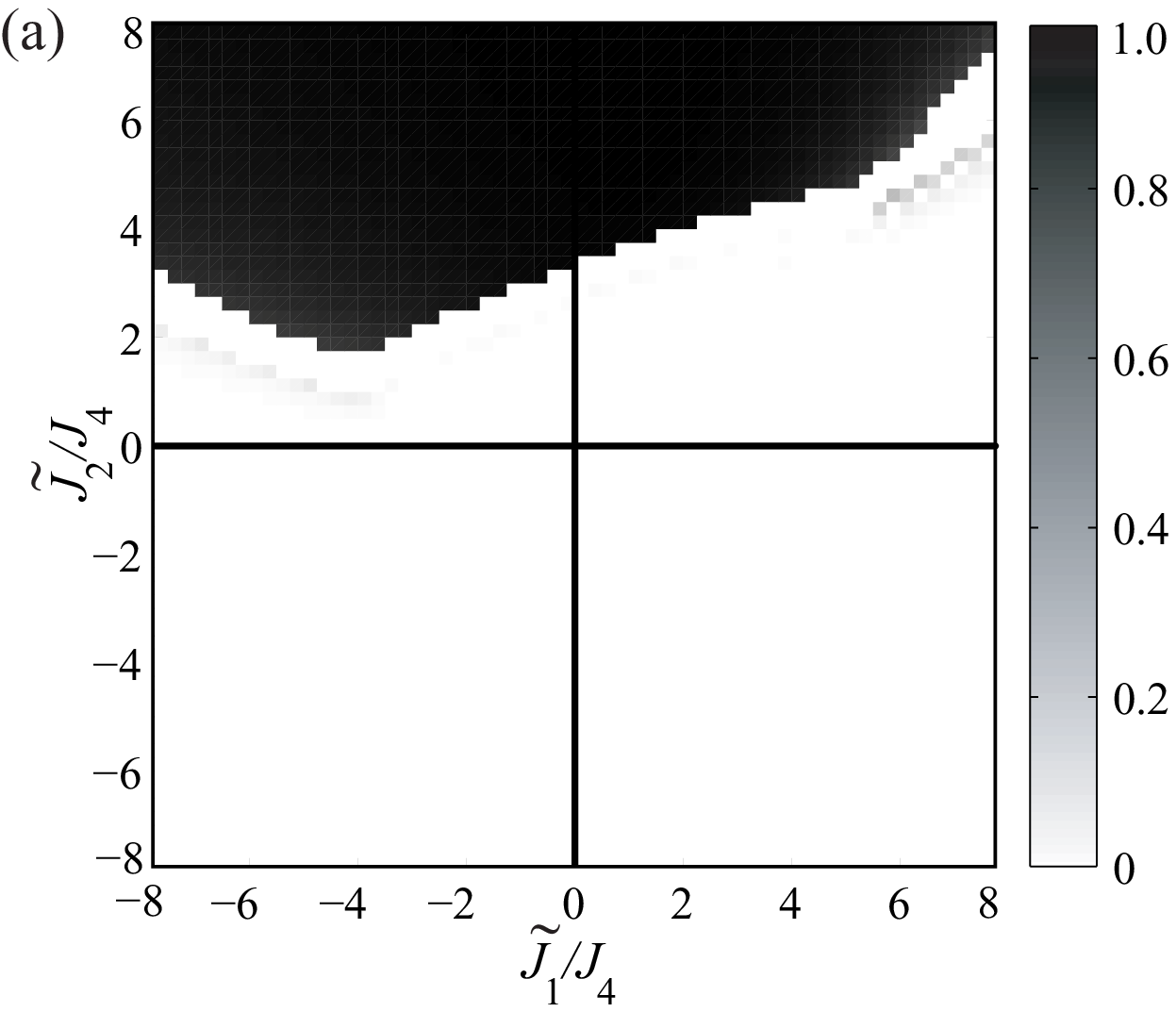}
  \includegraphics[height=6cm,clip]{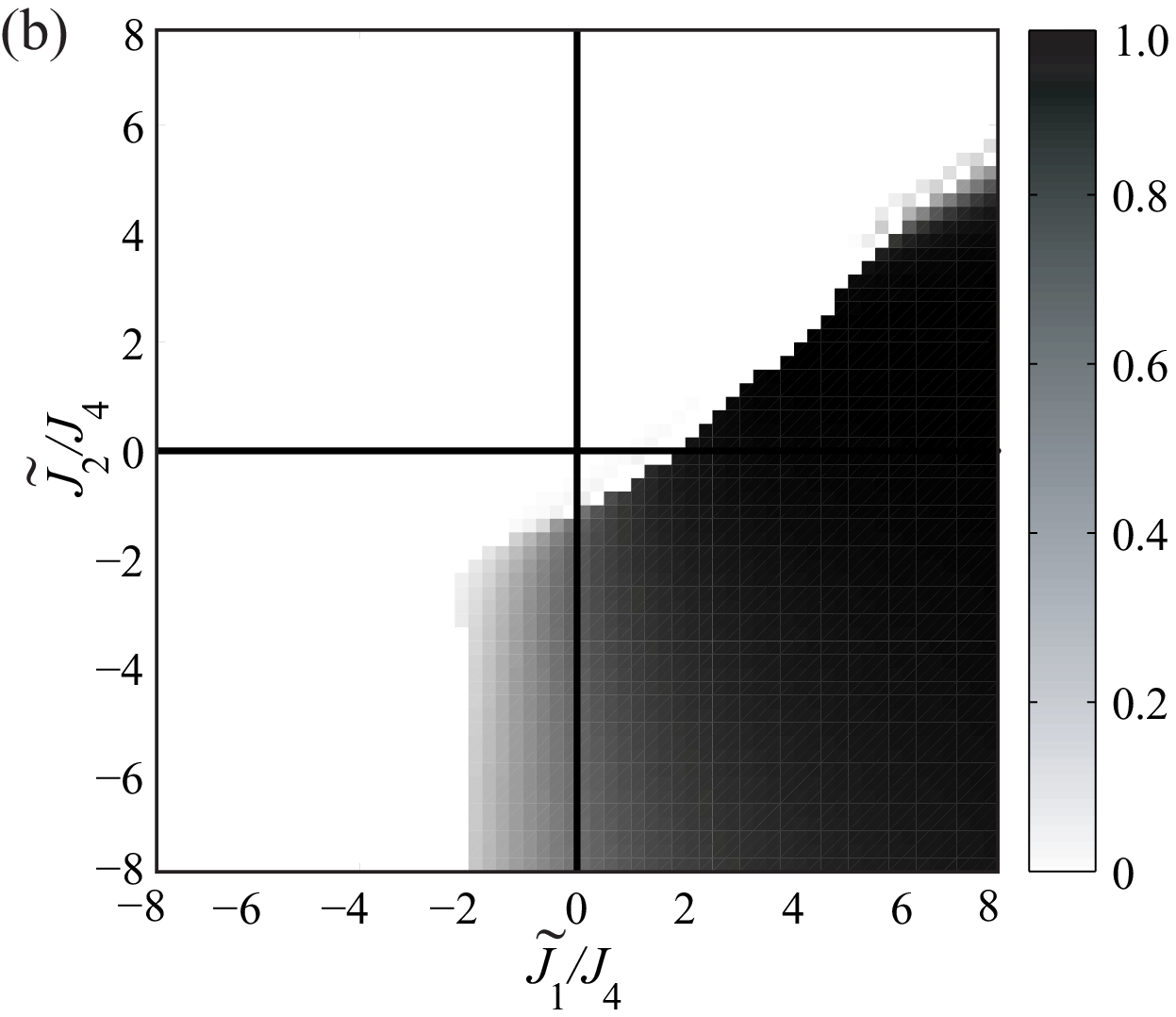}
  \caption{Overlaps of the ground state wavefunctions in the presence of
    the 4-particle ring exchange with the wavefunctions
    representing (a) the dimer and (b) the antiferromagnetic phase for
    $J_4=0$. The representative ground states (a) and (b) are obtained
    for $(\widetilde{J}_1, \widetilde{J}_2, J_4) = (1, 10, 0)$ and
    $(\widetilde{J}_1, \widetilde{J}_2, J_4) = (1,-10, 0)$,
    respectively.}
  \label{fig_6}
\end{figure}
At large values of $|\widetilde{J}_1|/J_4$ and
$|\widetilde{J}_2|/J_4$, one would expect to recover the phases
present in the absence of $J_4$. Thus, the large white area in
Fig.~\ref{fig_5} corresponding to total spin $S=0$ should contain the
antiferromagnetic phase, analogues of the dimer phases observed in the system without four-spin interactions, and possibly entirely new
phases as well. In order to distinguish between these phases, we first
calculate the overlap between the ground state wavefunctions in our
model and the ones representing the dimer and antiferromagnetic phases
in the well-studied model with $J_4=0$. The representative ground
state wavefunctions are obtained for the chain with $J_4 = 0$ and
typical parameter sets of $(\widetilde{J}_1, \widetilde{J}_2)$ chosen
deep in the dimer and antiferromagnetic phases of the phase diagram
shown in Fig.~\ref{fig_2}. The results for the chain with $N = 24$
sites are shown in Fig.~\ref{fig_6}. As can be seen from the figure,
the ground states for a broad region of large positive
$\widetilde{J}_2/J_4$ have a significant overlap with the
representative ground state of the dimer phase while the ground states
for large positive $\widetilde{J}_1/J_4$ and/or negative
$\widetilde{J}_2/J_4$ resemble very much the one belonging to the
antiferromagnetic phase. This behavior indicates the appearance of the
expected dimer and antiferromagnetic phases for large effective
couplings $|\widetilde{J}_1|/J_4$ and $|\widetilde{J}_2|/J_4$. We have confirmed the existence of these phases in the corresponding parameter regimes by studying the associated structure factors.

\begin{figure}[!t]
  \includegraphics[height=6.4cm,clip]{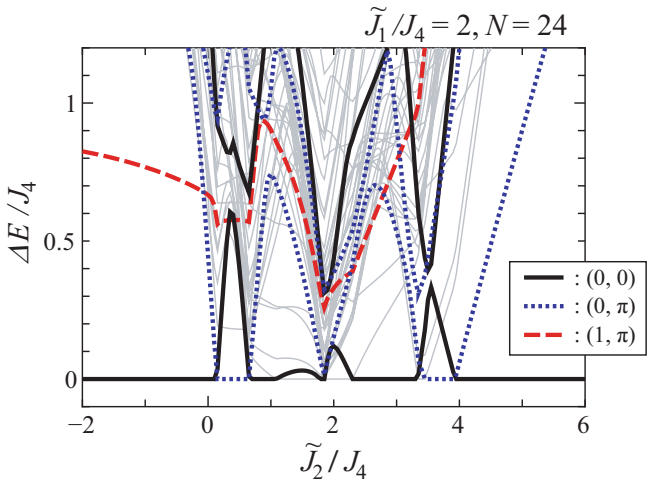}
  \caption{Excitation energies $\Delta E_n(S, Q)$ in the system of
    $N=24$ sites for $\widetilde{J}_1/J_4=2$ as functions of
    $\widetilde{J}_2/J_4$. The two-lowest levels are plotted for the
    subspaces of $(S,Q)=(0,0)$ and $(0,\pi)$ while only the lowest one
    is shown for all other subspaces. The energies for $(S,Q)=(0,0)$,
    $(0,\pi)$, and $(1,\pi)$ are plotted by thick solid, dotted,
    and dashed curves, respectively. The energies of the levels
    belonging to other subspaces are shown by thin gray curves.}
  \label{fig_7}
\end{figure}
In order to study and clarify the properties of the system in more
detail, we have calculated the excitation energies
\begin{equation}
  \Delta E_n(S, Q) = E_n (S, Q) - E_{\rm gs},
  \label{eq:gap}
\end{equation}
where $E_n(S, Q)$ is the energy of $n$-th lowest level in the subspace
characterized by the total spin $S$ and the momentum $Q$, and $E_{\rm gs}$ is the ground state energy. Figure \ref{fig_7} shows the results for the system of size $N = 24$,
obtained along the vertical line in the phase diagram given by $\widetilde{J}_1/J_4 = 2$. At large positive $\widetilde{J}_2/J_4$, the ground and first-excited states belong to the subspace $(S, Q) = (0, 0)$ and $(0, \pi)$, respectively.\cite{dimerdoublet-footnote} These states are expected to form the ground state doublet of the dimer phase in the thermodynamic limit. For
$\widetilde{J}_2/J_4>(\widetilde{J}_2/J_4)_{c,{\rm dim}}\sim 3.5$, one of the dimer doublet states is the ground state and the system is in the dimer phase. At smaller $\widetilde{J}_2/J_4$, both states of the dimer doublet shift upward and move away from the low-energy regime, while other states decrease steeply in energy and eventually become the ground state. We therefore take the point $(\widetilde{J}_2/J_4)_{c,{\rm dim}}$ as the boundary of the dimer phase. After the transition, the system enters a region with exotic ground states and a large number of low-lying excitations. We have numerically checked that these exotic ground states have no or, at most, negligibly small overlap with the ground state of either the dimer or antiferromagnetic phases. When $\widetilde{J}_2/J_4$ decreases further, the exotic states leave the low-energy regime and the system predictably enters the antiferromagnetic phase, which occurs for
$\widetilde{J}_2/J_4<(\widetilde{J}_2/J_4)_{c,{\rm AF}} \sim 0.1$.

\begin{figure}[!t]
  \includegraphics[height=7.0cm,clip]{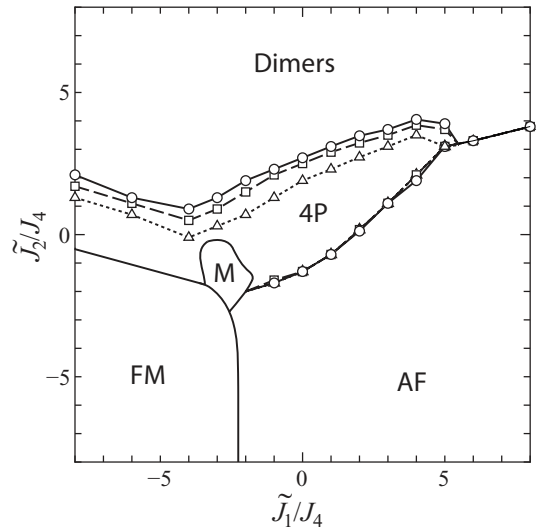}
  \caption{ The phase diagram of the Heisenberg chain including
    nearest neighbor, next-nearest neighbor, and 4-particle ring exchanges. The expected phases consist of a ferromagnetic and an antiferromagnetic phase as well as a dimer phase. In addition, a novel region ($4P$) dominated by the 4-particle ring exchange appears. The latter includes a phase of partial spin polarization ($M$). Triangles, squares and circles correspond to the boundaries obtained for $N= 16$, 20, and 24 sites, respectively. We note that although the phase of partial spin polarization persists as the system size is increased, its boundary with the 4P phase has a rather irregular size dependence and is represented approximately in the figure.} \label{fig_8}
\end{figure}

Performing the same type of analysis for several parameter lines, we
can estimate the phase boundaries $(\widetilde{J}_2/J_4)_{c, {\rm
    dim}}$ and $(\widetilde{J}_2/J_4)_{c, {\rm AF}}$ as functions of
$\widetilde{J}_1/J_4$. In the limit of large negative coupling
$\widetilde{J}_1/J_4 \to -\infty$, the boundary of the dimer phase
$(\widetilde{J}_2/J_4)_{c, {\rm dim}}$ approaches the line
$\widetilde{J}_1 = -0.38 \widetilde{J}_2$, suggesting a smooth
connection to the behavior for $\widetilde{J}_1 < 0$ and $J_4
= 0$ (cf. Ref.~\onlinecite{Chubukov}). In a similar fashion, at large
positive coupling $\widetilde{J}_1/J_4$, we find no indication for the
appearance of exotic phases after $\widetilde{J}_1/J_4 \ge 6$; the
data of the energy spectrum and the wavefunction overlaps show
essentially the same behaviors as those at $\widetilde{J}_1/J_4 \to
\infty$. We therefore conclude that there occurs a direct transition
between the dimer and antiferromagnetic phases and estimate the
transition line using the method of level spectroscopy, according to
which the transition point is determined by the level crossing between
the first-excited states in the dimer and antiferromagnetic
phases.\cite{Okamoto}

Combining all these phase boundaries and including the boundaries of the ferromagnetic and partially spin polarized phases which were obtained using the total spin of the ground state as a criterion, we determine the phase diagram in the $\widetilde{J}_1/J_4$ versus $\widetilde{J}_2/J_4$ plane. The result is shown in Fig.~\ref{fig_8}.  The phase diagram has similarities to the one obtained without the four-spin interaction term, see Fig.~\ref{fig_2}.  In particular, the expected ferromagnetic, antiferromagnetic, and dimer phases appear for large values of the effective couplings, $|\widetilde{J}_1|/J_4$ and $|\widetilde{J}_2|/J_4$. But more importantly, at not too large values of the effective couplings, new phases appear as a direct result of the new interaction term. We can identify a phase with partial spin polarization and a region occupied by one or several novel phases with total spin $S=0$. In the region where $J_4$ dominates, the ground state has no similarity at the level of wavefunctions with that of the conventional phases. It is important to note that the region occupied by the new phases becomes broader as the system size $N$ grows, indicating that it survives even in the thermodynamic limit. From the analysis of the wavefunction overlaps between the ground states, there are strong indications that the novel unpolarized region might consist of several different phases. Unfortunately, it has proven difficult to clarify the nature of the new phases and, in particular, discover the order parameters that characterize them based solely on the analysis of small systems. Therefore, the issue is relegated to future studies. In the absence of detailed understanding of its properties, we collectively dub the region of the phase diagram the ``4P'' phase.

\section{Phase diagram for realistic quantum wires}
\label{PDS} Having identified possible phases of the zigzag chain, the
most interesting question is which of the various phases appearing in the phase diagram Fig.~\ref{fig_8} are accessible in quantum wires.  At finite $r_\Omega$, the calculations of the exchange constants discussed in Sec.~\ref{EX} have to be completed in an important way by computing the prefactors $J_{l}$ in Eq.~(\ref{exch}). To that effect it is necessary to take into account Gaussian fluctuations around the classical exchange path.  We employ the method introduced by Voelker and Chakravarty\cite{Voelker} which, for the sake of completeness, is outlined in the Appendix \ref{appendix}. The prefactors have the form
\begin{equation}
  J^{*}_{l}=\frac{e^2}{\epsilon
    a_B}\,A_lF_l\,{r_\Omega}^{-\frac{5}{4}}\sqrt{\frac{\eta_l}{2\pi}},
  \label{ppf}
\end{equation}
where $F_l$ is density dependent. The factor $A_l$ is used
to account for multiple classical trajectories corresponding to the
same exchange process (see Appendix).  Table~\ref{tbl_1} contains the
values of $F_l$ we calculated for the various exchanges considered in
this work. Note that, in order to achieve a comparable level of
convergence, a more accurate determination of the instanton
trajectories was required for the calculation of the prefactors
$J^{*}_{l}$ than for the calculation of the exponents $\eta_l$. By
including up to 28 moving spectators on either side of the exchanging
particles, we have been able to achieve an accuracy better than 2\%.
\begin{table}[!t]
  \begin{ruledtabular}
    \begin{tabular}{ccccc}
      $\nu$ & $F_{1}$ & $F_{2}$ & $F_{3}$ & $F_{4}$\\\hline
      1.0  & 1.12 & $\simeq 6$ & 1.22 & 2.44\\
      1.1  & 1.04 & $\simeq 4$ & 1.03 & 1.73\\
      1.2  & 1.05 & 2.38       & 0.97 & 1.28\\
      1.3  & 1.08 & 1.86       & 0.97 & 1.15\\
      1.4  & 1.19 & 1.71       & 1.02 & 1.13\\
      1.5  & 1.40 & 1.63       & 1.14 & 1.18\\
      1.6  & 1.80 & 1.51       & 1.26 & 1.19\\
      1.7  & 2.07 & 1.07       & 0.81 & 0.50\\
    \end{tabular}
  \end{ruledtabular}
  \caption{\label{tbl_1}The numerically calculated values of the
    density dependent part $F_{l}$ of the exchange energy prefactor
    $J^{*}_{l}$, see Eq.~(\ref{ppf}), calculated with mobile
    spectators. For all the numbers reported, the accuracy is better
    than 2\%, except for $F_2$ at $\nu=1.0$, $1.1$, for which
    extrapolated values with an estimated error of $\sim 10\%$ are
    shown.}
\end{table}

\begin{figure}[!t]
  \includegraphics[height=6.0cm,clip]{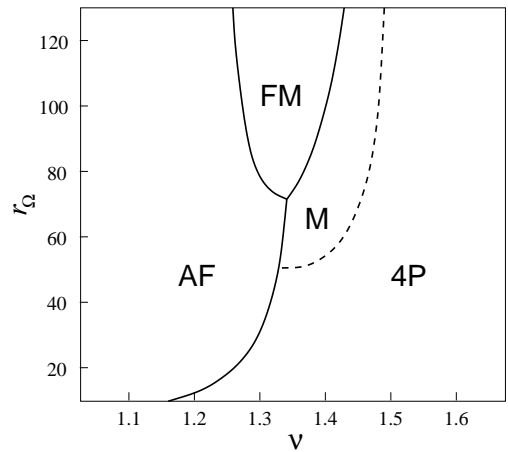}
  \caption{The phase diagram as a function of the dimensionless density $\nu$ and interaction strength $r_\Omega$. The various phases were obtained by first calculating the effective couplings $\widetilde{J}_1/J_4$ and $\widetilde{J}_2/J_4$ for a given point; subsequently, the corresponding phase was determined utilizing the calculated boundaries shown in Fig.~\ref{fig_8} for a system of $N=24$ sites.}
  \label{fig_9}
\end{figure}

We are now in a position to map out the areas of the phase diagram of
Fig.~\ref{fig_8} that are encountered as one traverses the density
region of interest for a given $r_\Omega$. The resulting phase diagram
obtained with the calculated exchange energies is shown in
Fig.~\ref{fig_9}.  Since the semiclassical approximation is applicable
only at $r_\Omega\gg1$, we do not extend the phase diagram to values
of $r_\Omega<10$.  It turns out that the spin polarized phases are
only realized at $r_\Omega\gtrsim 50$. On the other hand, the novel
``4P'' phase is expected to appear in a certain density range as long
as $r_\Omega\gg1$.

\section{Discussion}
\label{disc}
In the preceding sections we have studied the coupling of spins of
electrons forming a zigzag Wigner crystal in a parabolic confining
potential.  We have found that apart from the 2-particle exchange couplings between the nearest and next-nearest neighbor spins, the 3- and 4-particle ring exchange processes have to be taken into account.  At relatively low electron densities, when the transverse displacement of electrons is small compared to the distance between particles,
Fig.~\ref{fig_1}(b), the nearest-neighbor 2-particle exchange
dominates.  In this regime the spins form an antiferromagnetic ground
state, with low-energy excitations described by the Tomonaga-Luttinger
theory.  At relatively high densities, when the transverse
displacements are large, Fig.~\ref{fig_1}(c), the 4-particle ring
exchange processes dominate.  Since the ring exchange processes
involving even numbers of particles favor spin-unpolarized states, the
ground state of the system in this regime has zero total spin. Finally, if the confining potential is sufficiently shallow, so that the parameter $r_\Omega\gtrsim 50$, there is an intermediate density range in which the 3-particle exchange processes are important, and the ground state is spontaneously spin-polarized.  These results are summarized in Fig.~\ref{fig_9}.

We expect that the zigzag Wigner crystal state can be realized in
quantum wires.  In order for the zigzag crystal to form the confining
potential of the wire should be rather shallow, so that large values
$r_\Omega\gg 1$ of the parameter (\ref{eq:r_Omega}) could be achieved.
The exact shape of the confining potential in existing wires is not
well known.  Using the quoted value of subband spacing $\sim 20$~meV
we estimate that the parameter $r_\Omega$ is of order unity in
cleaved-edge-overgrowth wires.\cite{yacoby} The confining potential in
split-gate quantum wires tends to be more shallow.  For a typical
value 1~meV of subband spacing we estimate $r_\Omega\approx 6$.
Finally, for \textit{p}-type quantum wires\cite{klochan,daneshvar}
with subband spacing $\sim300~\mu$eV we estimate $r_\Omega\approx 20$.
These hole systems are the most promising devices for observation of
the zigzag Wigner crystal.

Given the relatively modest values of $r_\Omega\lesssim 20$ in the
existing quantum wire structures, we do not expect that the
spontaneously spin-polarized ground state will be easily observed in
experiments. Instead, we expect that as the density of charge carriers
is increased, a transition from antiferromagnetism to a state
dominated by 4-particle ring exchanges will occur.  We have found that the ground state in this phase has a complicated size dependence, which makes it very difficult to identify its nature by exact diagonalization of finite-size chains.  To fully understand the spin properties in the
high density regime, further studies of zigzag ladders with ring
exchange coupling are needed.

\acknowledgments
We acknowledge helpful discussions with A. L{\"a}uchli and T. Momoi. This work was supported by the U. S. Department of Energy, Office of Science, under Contract No.~DE-AC02-06CH11357.  T.H. was supported in part by a Grant-in-Aid from the Ministry of Education, Culture, Sports, Science and Technology (MEXT) of Japan (Grant Nos.~16740213 and 18043003). Part of the calculations were performed at the Ohio Supercomputer Center thanks to a grant of computing time.

\appendix*
\section{Calculation of the prefactors}
\label{appendix} In order to find the prefactors $J_l^*$ in the
expressions for the exchange constants, fluctuations around the
instanton trajectory have to be taken into account.  The Euclidean
(imaginary time) path integral for the propagator $G({\bf R}_1,{\bf
  R}_2;T)=\langle{\bf R}_{1}|e^{-T H}|{\bf R}_{2}\rangle$ can be
written as
\begin{equation}
  G({\bf R}_1,{\bf R}_2;T)=\int_{{\bf R}(0)={\bf R}_1}^{{\bf
      R}(T)={\bf R}_2}\!\!\!\mathcal{D}{\bf R}\,e^{-\frac{1}{\hbar}S[{\bf
      R}]},
\end{equation}
where the Euclidean action is given by
\begin{equation}
  S[{\bf
    R}] = \int_{0}^{T}\!\!\!\mathrm{d}\tau
  \left\{\frac{m}{2}\left(\frac{\mathrm{d}{\bf
          R}}{\mathrm{d}\tau}\right)^2+V({\bf R})\right\}.
\end{equation}
Here ${\bf R}$ is a $M$-dimensional position vector, where $M/2$ is
the total number of moving particles, including the exchanging
particles as well as the spectators.  In the semiclassical limit, the
integral is dominated by the classical path ${\bf R}_{cl}(\tau)$ that
extremizes the action $S$ for a given exchange process. (The exponents
$\eta$ are given as $\eta=S[{\bf R}_{cl}]/(\hbar\sqrt{r_\Omega})$.)
The Gaussian quantum fluctuations about the classical path can be
taken into account by defining fluctuation coordinates ${\bf
  u}(\tau)\equiv{\bf R}(\tau)-{\bf R}_{cl}(\tau)$ and subsequently
expanding the action to second order. We obtain for the propagator
\begin{align}
  &G({\bf R}_1,{\bf R}_2;T)=F[{\bf R}_{cl}]e^{-\frac{1}{\hbar}S[{\bf
      R}_{cl}]},\\\notag\\
  &F[{\bf R}_{cl}]=\int_{{\bf u}(0)=0}^{{\bf
      u}(T)=0}\!\!\!\mathcal{D}{\bf
    u}(\tau)\,e^{-\frac{1}{\hbar}\delta S[{\bf
      u}(\tau)]},\label{SF_1}\\\notag\\
  &\delta S[{\bf u}(\tau)] = \frac{m}{2} \int_{0}^{T}\!\!\mathrm{d}\tau
  \left\{\dot{{\bf u}}^2(\tau)+{\bf u}^T(\tau)\mathcal{H}(\tau){\bf
      u}(\tau)\right\}, \label{SF_2}\\\notag\\
  &\mathcal{H}_{kp}(\tau)=\frac{1}{m}\left.\frac{\partial^2V({\bf
        R})}{\partial R_{k}\partial R_{p}}\right|_{{\bf R}={\bf
      R}_{cl}(\tau)}.
\end{align}
In the preceding formulas, ${\bf R}_1$ and ${\bf R}_2$ correspond to
two configurations of electrons that minimize the electrostatic
potential $V({\bf R})$ describing electron-electron interactions as
well as the confining potential. The exchange constant is related to
the ratio of the propagator for a particular exchange process ${\bf
  R}_1\rightarrow{\bf R}_2$, divided by the propagator for the trivial
path ${\bf R}_{cl}(\tau)={\bf R}_1$:
\begin{equation}
  G=\frac{F[{\bf R}_{cl}]}{F[{\bf R}_{1}]}\,e^{-\frac{1}{\hbar}S[{\bf
      R}_{cl}]}.
\end{equation}
We start from the expression for the propagator in the semiclassical
limit and proceed by partitioning the time interval $[0,T]$ into $N$
subintervals $(\tau_0,\tau_1)$, $(\tau_1,\tau_2)$, $\dots$,
$(\tau_{N-1},\tau_N)$, with $\tau_0=0$, $\tau_N=T$. The partition is chosen sufficiently fine as to enable the approximation that in each
subinterval, the Hessian matrix $\mathcal{H}(\tau)$ of the second
derivative of the potential can be considered time independent,
$\mathcal{H}(\tau)\simeq \mathcal{H}(\tau_{\nu})\equiv
\mathcal{H}^{\nu}$. (In what follows, we use the convention that for
the fluctuation coordinates, superscripts denote time subinterval,
while subscripts denote spatial coordinate.) Subsequently the path
integral is calculated as a product of path integrals over the
partitioned interval. Moreover, each individual path integral is that
of a multidimensional harmonic oscillator, for which analytic results
exist. We then have
\begin{widetext}
  \begin{equation}
    F[{\bf R}_{cl}]=\int\!\!\mathrm{d}{\bf u}^1\,G_1({\bf u}^1,{\bf
      u}^0;\tau_1-\tau_0)\dots\int\!\!\mathrm{d}{\bf
      u}^{N-1}\,G_{N-1}({\bf u}^{N-1},{\bf
      u}^{N-2};\tau_{N-1}-\tau_{N-2})\, G_N({\bf u}^N,{\bf
      u}^{N-1};\tau_N-\tau_{N-1}),
  \end{equation}
  and the propagator for each subinterval is
  \begin{equation}
    G_{\nu}({\bf u}^{\nu},{\bf
      u}^{\nu-1};\tau_{\nu}-\tau_{\nu-1})=\int_{{\bf u}(\tau_{\nu-1})={\bf
        u}^{\nu-1}}^{{\bf u}(\tau_\nu)={\bf u}^{\nu}}\!\!\!\mathcal{D}{\bf
      u}(\tau)
    \,\exp \left\{-\frac{m}{2\hbar}
      \int_{\tau_{\nu-1}}^{\tau_{\nu}}\!\!\mathrm{d}\tau \left[\dot{{\bf
            u}}^2(\tau)+{\bf u}^T(\tau)\mathcal{H}^{\nu}{\bf
          u}(\tau)\right]\right\}.
  \end{equation}
\end{widetext}
Within each imaginary time subinterval, we define orthonormal
eigenvectors $q_{\mu}^{\nu}=\sum_{k=1}^{M}U^{\nu}_{k
  \mu}u_{k}^{\nu}$. The unitary matrix $U^{\nu}$ is such that
$\mathcal{H}^{\nu}=U^{\nu}\Lambda^{\nu}(U^{\nu})^T$, with $\Lambda$ a
diagonal matrix of eigenvalues $(\lambda^{\nu}_{\mu})^2$, $\mu=1\dots
M$, where $M$ is the number of spatial coordinates. Then one
immediately obtains
\begin{widetext}
  \begin{equation}
    G_{\nu}({\bf q}^{\nu},{\bf
      q}^{\nu-1};\tau_{\nu}-\tau_{\nu-1})=\int_{{\bf q}(\tau_{\nu-1})={\bf
        q}^{\nu-1}}^{{\bf q}(\tau_\nu)={\bf q}^{\nu}}\!\!\!\mathcal{D}{\bf
      q}(\tau)
    \,\exp
    \left\{-\frac{m}{2\hbar}\int_{\tau_{\nu-1}}^{\tau_{\nu}}\!\!\mathrm{d}\tau
      \left[\dot{{\bf
            q}}^2(\tau) + {\bf q}^T(\tau)\Lambda^{\nu}{\bf
          q}(\tau)\right]\right\}
    =\bar{F}[{\bf q}_{cl}]e^{-\frac{1}{\hbar}\delta S[{\bf q}_{cl}]},
  \end{equation}
  where ${\bf q}_{cl}$ is the classical trajectory connecting ${\bf
    q}^{\nu-1}$ and ${\bf q}^\nu$. Considering the fluctuation part
  first, we obtain an elementary path integral
  \begin{equation}
    \bar{F}[{\bf q}_{cl}]=\int_{{\bf q}(\tau_{\nu-1})=0}^{{\bf
        q}(\tau_\nu)=0}\!\!\!\mathcal{D}{\bf
      q}(\tau)\, \exp\left\{-\frac{m}{2\hbar}
      \int_{\tau_{\nu-1}}^{\tau_{\nu}}\!\!\mathrm{d}\tau\,{\bf
        q}^T(\tau)\left[-\frac{\mathrm{d}^2}{\mathrm{d}\tau^2} +
        \Lambda^{\nu}\right]{\bf
        q}(\tau)\right\}=\prod_{\mu=1}^{M}\sqrt{\frac{B_{\mu}^{\nu}}{2\pi}},
  \end{equation}
\end{widetext}
where
\begin{equation}
  B_{\mu}^{\nu}=\frac{m\lambda_{\mu}^{\nu}}
  {\hbar\sinh(\lambda^{\nu}_{\mu}\Delta\tau_{\nu})},
\end{equation}
and $\Delta\tau_\nu=\tau_n-\tau_{n-1}$. The exponent $\delta S[{\bf q}_{cl}]$ can now be calculated explicitly
\begin{align}
  \frac{\delta S[{\bf q}_{cl}]}{\hbar}&=
  \frac{1}{2}\sum_{\mu=1}^{M}B_{\mu}^{\nu}
  \left\{[(q_{\mu}^{\nu})^{2}_{cl}+(q_{\mu}^{\nu-1})^{2}_{cl}]
    \cosh(\lambda^{\nu}_{\mu}\Delta\tau_{\nu})\right. \notag\\
  &\phantom{\frac{1}{2}\sum_{\mu=1}^{M}B_{\mu}^{\nu}\left\{\right\}}\left.
    -2(q_{\mu}^{\nu})_{cl}(q_{\mu}^{\nu-1})_{cl}\right\}.
\end{align}
The subscript ``$cl$'' used for notational clarity will be
subsequently dropped from all expressions. With some additional
algebra, the remaining integral is easily evaluated.  With the
following definitions
\begin{align}
  \Gamma_{kp}^{\nu}&=\sum_{\mu=1}^{M}U^{\nu}_{k \mu}
  \frac{m\lambda^{\nu}_{\mu}}{\hbar\tanh(\lambda^{\nu}_{\mu}\Delta\tau_{\nu})}
  U^{\nu}_{p \mu}\\
  \Delta_{kp}^{\nu}&=\sum_{\mu=1}^{M}U^{\nu}_{k
    \mu}\frac{m\lambda^{\nu}_{\mu}}
  {\hbar\sinh(\lambda^{\nu}_{\mu}\Delta\tau_{\nu})} U^{\nu}_{p \mu},
\end{align}
we find
\begin{equation}
  F[{\bf R}_{cl}]=(2\pi)^{-\frac{M}{2}}
  \prod_{\nu=1}^{N}\prod_{\mu=1}^{M}\sqrt{B_{\mu}^{\nu}}\frac{1}{\sqrt{\det\Omega}},
\end{equation}
where the $M(N-1)\times M(N-1)$ matrix $\Omega_{kp}^{\nu\lambda}$ has
components
\begin{equation}
  \Omega_{kp}^{\nu\lambda} = (\Gamma_{kp}^{\nu}+\Gamma_{kp}^{\nu+1})
  \delta^{\nu,\lambda} - \Delta_{kp}^{\nu}\delta^{\nu,(\lambda+1)} -
  \Delta_{kp}^{\lambda}\delta^{\nu,(\lambda-1)}.
\end{equation}
The calculation of $F[{\bf R}_{1}]$ is carried out in an identical
manner and the subscript ``0'' will be used to distinguish the results
pertaining to that calculation.

Finally, one has to account for the existence of an eigenvalue of the
matrix $\Omega$ which is identically zero in the continuum limit and
corresponds to the zero mode associated with uniform translation of
the instanton in imaginary time. The procedure is
standard\cite{Coleman} and we simply report the result for the
prefactor here. One obtains
\begin{equation}
  G=T\left(\frac{S}{2\pi\hbar m}\right)^{\frac{1}{2}}
  \displaystyle\prod_{\nu=1}^{N} \prod_{\mu=1}^{M}
  \sqrt{\frac{B_{\mu}^{\nu}}{B_{\mu,0}^{\nu}}}
  \sqrt{\frac{\det\Omega_0}{\det'\Omega}},
\end{equation}
where the primed determinant implies the exclusion of the eigenvalue
corresponding to the zero mode. Reverting to the system of units used
in this work, the prefactor of the exchange energy is given by
\begin{equation}
  J^{*}_{l}=\frac{e^2}{\epsilon a_B}\,A_l\, {r_\Omega}^{-\frac{5}{4}}
  \sqrt{\frac{\eta_l}{2\pi}}
  \displaystyle\prod_{\nu=1}^{N}\prod_{\mu=1}^{M}
  \sqrt{\frac{B_{\mu}^{\nu}}{B_{\mu,0}^{\nu}}}
  \sqrt{\frac{\det\Omega_0}{\det'\Omega}}.\label{J}
\end{equation}
The additional factor $A_l$ is used to account for multiple classical
trajectories corresponding to the same exchange process, as happens
for the case of nearest and next-nearest neighbor exchanges (i.e.,
$A_1=A_2=2$, whereas $A_l=1$ for $l\geq3$).

The numerical implementation of the method outlined above is
straightforward. In particular, the quantity that needs to be
numerically calculated, once for each type of exchange at all
densities of interest, is
\begin{equation}
  F_l=\displaystyle \prod_{\nu=1}^{N} \prod_{\mu=1}^{M}
  \sqrt{\frac{B_{\mu}^{\nu}}{B_{\mu,0}^{\nu}}}
  \sqrt{\frac{\det\Omega_0}{\det'\Omega}}. \label{F}
\end{equation}
We note here that the eigenvalue corresponding to the zero mode is
easily calculated with the same procedure used by Voelker and
Chakravarty\cite{Voelker}. In the definition of the prefactor, see Eqs.~(\ref{SF_1}) and (\ref{SF_2}), one replaces $\mathcal{H}(\tau)$
with $\mathcal{H}(\tau)-\lambda$, with $\lambda$ a free parameter. Subsequently, a numerical search for the smallest eigenvalue that results in $1/F(\lambda)=0$ is carried out. The
smallest eigenvalue corresponds to the zero mode, and for a finite
partition of the imaginary time interval it is a small but finite
number.


\begin{thebibliography}{99}

\bibitem{Thomas} K.~J.~Thomas, J.~T.~Nicholls, M.~Y.~Simmons,
  M.~Pepper, D.~R.~Mace, and D.~A.~Ritchie, Phys.\ Rev.\
  Lett. \textbf{77}, 135 (1996).

\bibitem{Kristensen_1} A.~Kristensen, J.~B.~Jensen, M.~Zaffalon,
  C.~B.~S\o rensen, S.~M.~Reimann, P.~E.~Lindelof, M.~Michel, and
  A.~Forchel, J.\ Appl.\ Phys. \textbf{83}, 607 (1998).

\bibitem{Kristensen} A.~Kristensen, H.~Bruus, A.~E.~Hansen,
  J.~B.~Jensen, P.~E.~Lindelof, C.~J.~Marckmann, J.~Nyg\aa rd,
  C.~B.~S\o rensen, F.~Beuscher, A.~Forchel, and M.~Michel, Phys.\
  Rev.\ B \textbf{62}, 10950 (2000).

\bibitem{Thomas_1} K.~J.~Thomas, J.~T.~Nicholls, N.~J.~Appleyard,
  M.~Y.~Simmons, M.~Pepper, D.~R.~Mace, W.~R.~Tribe, and
  D.~A.~Ritchie, Phys.\ Rev.\ B \textbf{58}, 4846 (1998).

\bibitem{Kane} B.~E.~Kane, G.~R.~Facer, A.~S.~Dzurak, N.~E.~Lumpkin,
  R.~G.~Clark, L.~N.~Pfeiffer, and K.~W.~West, Appl.\ Phys.\
  Lett. \textbf{72}, 3506 (1998).

\bibitem{Thomas_2} K.~J.~Thomas, J.~T.~Nicholls, M.~Pepper,
  W.~R.~Tribe, M.~Y.~Simmons, and D.~A.~Ritchie, Phys.\ Rev.\ B
  \textbf{61}, R13365 (2000).

\bibitem{Reilly_1} D.~J.~Reilly, G.~R.~Facer, A.~S.~Dzurak,
  B.~E.~Kane, R.~G.~Clark, P.~J.~Stiles, R.~G.~Clark, A.~R.~Hamilton,
  J.~L.~O'Brien, N.~E.~Lumpkin, L.~N.~Pfeiffer, and K.~W.~West, Phys.\
  Rev.\ B \textbf{63}, 121311(R) (2001).

\bibitem{Cronenwett} S.~M.~Cronenwett,
  H.~J.~Lynch, D.~Goldhaber-Gordon, L.~P.~Kouwenhoven, C.~M.~Marcus,
  K.~Hirose, N.~S.~Wingreen, and V.~Umansky, Phys.\ Rev.\
  Lett. \textbf{88}, 226805 (2002).

\bibitem{Reilly} D.~J.~Reilly, T.~M.~Buehler, J.~L.~O'Brien,
  A.~R.~Hamilton, A.~S.~Dzurak, R.~G.~Clark, B.~E.~Kane,
  L.~N.~Pfeiffer, and K.~W.~West, Phys.\ Rev.\ Lett. \textbf{89},
  246801 (2002).

\bibitem{Crook} R.~Crook, J.~Prance, K.~J.~Thomas, S.~J.~Chorley,
  I.~Farrer, D.~A.~Ritchie, M.~Pepper, and C.~G.~Smith, Science
  \textbf{312}, 1359 (2006).

\bibitem{depicciotto} R.~de Picciotto, L.~N.~Pfeiffer, K.~W.~Baldwin,
  and K.~W.~West, Phys.\ Rev.\ B \textbf{72}, 033319 (2005).

\bibitem{Danneau} R.~Danneau, W.~R.~Clarke, O.~Klochan,
  A.~P.~Micolich, A.~R.~Hamilton, M.~Y.~Simmons, M.~Pepper, and
  D.~A.~Ritchie, Appl.\ Phys.\ Lett. \textbf{88}, 012107 (2006).

\bibitem{klochan} O.~Klochan, W.~R.~Clarke, R.~Danneau,
  A.~P.~Micolich, L.~H.~Ho, A.~R.~Hamilton, K.~Muraki, and
  Y.~Hirayama, Appl.\ Phys.\ Lett. \textbf{89}, 092105 (2006).

\bibitem{Rokhinson} L.~P.~Rokhinson, L.~N.~Pfeiffer, and K.~W.~West,
  Phys.\ Rev.\ Lett. \textbf{96}, 156602 (2006).

\bibitem{Berggren} C.-K.~Wang and K.-F.~Berggren, Phys.\ Rev.\ B
  \textbf{54}, R14257 (1996); \textbf{57}, 4552 (1998);
  A.~A.~Starikov, I.~I.~Yakimenko, and K.-F.~Berggren, Phys.\ Rev.\ B
  \textbf{67}, 235319 (2003).

\bibitem{Spivak} B.~Spivak and F.~Zhou, Phys.\ Rev.\ B \textbf{61},
  16730 (2000).

\bibitem{flambaum} V.~V.~Flambaum and M.~Yu.~Kuchiev, Phys.\ Rev.\ B
  {\bf 61}, R7869 (2000).

\bibitem{rejec} T.~Rejec, A.~Ram\u{s}ak, and J.~H.~Jefferson, Phys.\
  Rev.\ B \textbf{62}, 12985 (2000).

\bibitem{bruus} H.~Bruus, V.~V.~Cheianov, and K.~Flensberg, Physica E
  \textbf{10}, 97 (2001).

\bibitem{hirose} K.~Hirose, S.~S.~Li, and N.~S.~Wingreen, Phys.\ Rev.\
  B \textbf{63}, 033315 (2001).

\bibitem{sushkov} O.~P.~Sushkov, Phys.\ Rev.\ B \textbf{64}, 155319
  (2001); Phys.\ Rev.\ B \textbf{67}, 195318 (2003).

\bibitem{Meir} Y.~Meir, K.~Hirose, and N.~S.~Wingreen, Phys.\ Rev.\
  Lett. \textbf{89}, 196802 (2002).

\bibitem{Tokura} Y.~Tokura and A.~Khaetskii, Physica E \textbf{12},
  711 (2002).

\bibitem{Matveev} K.~A. Matveev, Phys.\ Rev.\ Lett. \textbf{92},
  106801 (2004); Phys.\ Rev.\ B \textbf{70}, 245319 (2004).

\bibitem{Meir2} T.~Rejec and Y.~Meir, Nature \textbf{442}, 900 (2006).

\bibitem{bruus1} H.~Bruus and K.~Flensberg, Semicond.\ Sci.\ Technol.
  \textbf{13}, A30 (1998).

\bibitem{seelig} G.~Seelig and K.~A.~Matveev, Phys.\ Rev.\ Lett.
  \textbf{90}, 176804 (2003).

\bibitem{Lieb} E.~Lieb and D.~Mattis, Phys.\ Rev. \textbf{125}, 164
  (1962).

\bibitem{us} A.~D.~Klironomos, J.~S.~Meyer and K.~A.~Matveev,
  Europhys.\ Lett.\ \textbf{74}, 679 (2006).

\bibitem{Schulz} H.~J.~Schulz, Phys.\ Rev.\ Lett. \textbf{71}, 1864
  (1993).

\bibitem{hasse} R.~W.~Hasse and J.~P.~Schiffer, Ann.\
  Phys. \textbf{203}, 419 (1990).

\bibitem{Piacente} G.~Piacente, I.~V.~Schweigert, J.~J.~Betouras, and
  F.~M.~Peeters, Phys.\ Rev.\ B \textbf{69}, 045324 (2004).

\bibitem{hausler} W.~H\"ausler, Z.\ Phys.\ B \textbf{99}, 551 (1996).

\bibitem{KRM} A.~D.~Klironomos, R.~R.~Ramazashvili, and K.~A.~Matveev,
  Phys.\ Rev.\ B \textbf{72}, 195343 (2005).

\bibitem{Fogler} M.~M.~Fogler and E.~Pivovarov, Phys.\ Rev.\ B
  \textbf{72}, 195344 (2005); J.\ Phys.: Condens.\ Matter {\bf 18}, L7 (2006).

\bibitem{Roger} M.~Roger, Phys.\ Rev.\ B \textbf{30}, 6432 (1984).

\bibitem{Katano} M.~Katano and D.~S.~Hirashima, Phys.\ Rev.\ B
  \textbf{62}, 2573 (2000).

\bibitem{Voelker} K.~Voelker and S.~Chakravarty, Phys.\ Rev.\ B
  \textbf{64}, 235125 (2001).

\bibitem{Bernu} B.~Bernu, L.~Candido, and D.~M.~Ceperley, Phys.\ Rev.\
  Lett. \textbf{86}, 870 (2001).

\bibitem{Thouless} D.~J.~Thouless, Proc.\ Phys.\ Soc.\ London
  \textbf{86}, 893 (1965).

\bibitem{Majumdar} C.~K.~Majumdar and D.~K.~Ghosh, J.\ Math.\
  Phys. \textbf{10}, 1388 (1969); \textbf{10}, 1399 (1969).

\bibitem{Haldane} F.~D.~M.~Haldane, Phys.\ Rev.\ B \textbf{25}, R4925
  (1982).

\bibitem{Okamoto} K.~Okamoto and K.~Nomura, Phys.\ Lett.\ A
  \textbf{169}, 433 (1992).

\bibitem{Eggert} S.~Eggert, Phys.\ Rev.\ B \textbf{54}, R9612 (1996).

\bibitem{White} S.~R.~White and I.~Affleck, Phys.\ Rev.\ B
  \textbf{54}, 9862 (1996).

\bibitem{Hamada} T.~Hamada, J.~Kane, S.~Nakagawa, and Y.~Natsume, J.\
  Phys.\ Soc.\ Jpn. \textbf{57}, 1891 (1988).

\bibitem{Tonegawa} T.~Tonegawa and I.~Harada, J.\ Phys.\ Soc.\ Jpn.
  \textbf{58}, 2902 (1989).

\bibitem{Chubukov} A.\ V.\ Chubukov, Phys.\ Rev.\ B \textbf{44}, 4693
  (1991).

\bibitem{Allen} D.~Allen, F.~H.~L.~Essler, and A.~A.~Nersesyan, Phys.\
  Rev.\ B \textbf{61}, 8871 (2000).

\bibitem{Itoi} C.~Itoi and S.~Qin, Phys.\ Rev.\ B \textbf{63}, 224423
  (2001).

\bibitem{Muramoto} N.~Muramoto and M.~Takahashi, J.\ Phys.\ Soc.\
  Jpn. \textbf{68}, 2098 (1999).

\bibitem{dimerdoublet-footnote} To be precise, we have found that the
  ground state at large $\widetilde{J}_2/J_4$ belongs to the subspace
  $(S, Q) = (0, 0)$ [$(0, \pi)$] for $N = 8m$ [$8m+4$], where $m$ is
  an integer, while the first-excited state belongs to the subspace
  $(S, Q) = (0, \pi)$ [$(0, 0)$].

\bibitem{yacoby} A.~Yacoby, H.~L.~Stormer, N.~S.~Wingreen,
  L.~N.~Pfeiffer, K.~W.~Baldwin, and K.~W.~West, Phys.\ Rev.\
  Lett. \textbf{77}, 4612 (1996).

\bibitem{daneshvar} A.~J.~Daneshvar, C.~J.~B.~Ford, A.~R.~Hamilton,
  M.~Y.~Simmons, M.~Pepper, and D.~A.~Ritchie, Phys. Rev. B
  \textbf{55}, R13409 (1997).

\bibitem{Coleman} S.~Coleman, \textit{Aspects of Symmetry} (Cambridge
  University Press, New York, 1988).

\end{thebibliography}
\end{document}